\documentclass[pra,preprint,tightenlines,showpacs,twocolumn,10pt]{revtex4} 
\usepackage{epsfig} 

\begin{document} 

\title{ Inelastic collisions of relativistic 
electrons with atomic targets assisted by a laser field } 
                                      
\author{ A.B.Voitkiv, B.Najjari and J.Ullrich} 
\affiliation{ Max-Planck-Institut f\"ur Kernphysik, 
Saupfercheckweg 1, D-69117 Heidelberg, Germany } 

\date{\today} 

\begin{abstract} 

We consider inelastic collisions between 
relativistic electrons and atomic targets 
assisted by a low-frequency laser field 
in the case when this field is still much weaker than 
the typical internal fields in the target. 
Concentrating on target transitions 
we show that they can be substantially 
affected by the presence of the laser field.   
This may occur either via strong modifications in  
the motion of the relativistic electrons 
caused by the electron-laser interaction  
or via the Compton effect when the incident 
electrons convert laser photon(s) into 
photons with frequencies equal to 
target transition frequencies.  

\end{abstract}


\maketitle 


The theoretical studies of 
electron-atom collisions in the presence 
of an intense laser field have already 
quite a long history. 
Starting with the sixties,   
the different aspects of scattering of 
a non-relativistic electron moving 
in a laser field on an external potential  
had been considered in numerous papers,   
later on these studies were extended 
to field-assisted inelastic electron-atom collisions 
in which atomic internal degrees of freedom 
are excited (for a review see \cite{ehl}).   

More recently much attention has been paid to 
processes involving relativistic electrons 
in strong laser fields 
(for a recent review see \cite{ehl-new}) 
including field-assisted collisions between 
relativistic electrons and atomic targets 
(see \cite{kam-ehl}-\cite{miloc-ehl} and references therein).  
To our knowledge, however, 
in all the studies of such collisions    
the target was regarded merely 
as a source of an external potential acting 
on the electrons whose internal structure 
is not influenced by the collision process. 
In this communication we report first 
theoretical results for 
field-assisted collisions of relativistic electrons 
with atomic (ionic) targets which explicitly 
take into account the internal degrees 
of freedom of the target. 

Inelastic laser-field-free collisions of relativistic 
electrons with atoms have been extensively studied 
in the past (see for a review \cite{e-atom-review}). 
The addition of a laser field into such collisions 
could substantially modify the states of the electron and/or 
the atom that in general may lead to various effects.   
For instance, if the laser field is strong enough 
it can, even without any collision, 
cause ionization of an atomic target. 
In what follows, however, 
we shall focus on a different situation. 
Namely, one can choose the parameters 
of the laser field in such a way that 
the field {\it per se} 
does not affect the target directly  
but can very effectively interact with  
the incident electrons. Then, 
solely via this interaction, 
the field may modify inelastic target  
cross sections compared to the case of 
the field-free collisions. 

Atomic units are used throughout 
unless otherwise stated. 

Let at the remote past ($t \to -\infty$) 
and future ($t \to +\infty$),   
when the interactions between the projectile-electron and 
the atomic target and the laser field are supposed 
not to occur, the states of this electron 
are given, respectively, by 
\begin{eqnarray} 
\Psi_i(x) &=& \frac{1}{V_e} %
\sqrt{ \frac{mc^2}{E_i} } u_i(p,s)  %
e^{-i p_{i}^{\nu} x_{\nu}} 
\nonumber \\ 
\Psi_f(x) &=& \frac{1}{V_e} %
\sqrt{ \frac{mc^2}{E_f} } u_fi(p,s) %
e^{-i p_{f}^{\nu} x_{\nu}},  
\label{electron-1}
\end{eqnarray} 
where $x^{\nu}=(ct; {\bf x})$, 
$p_{\mu}=(E/c; - {\bf P})$ and $s$  
are the electron's space-time coordinates, 
four-momentum and spin, respectively,  
$u$ is the constant four-spinor  
for a free electron,   
$V_e$ is the normalization volume 
and $c$ the speed of light. 
In (\ref{electron-1}) 
and below the summation over repeated 
Greek indices is implied. 
    
Let $\psi_i$ and $\psi_f$ be the initial and final 
(undistorted) internal states of the atomic 
(ionic) target with the corresponding energies given 
by $\epsilon_i$ and $\epsilon_f$, respectively.   

The laser field is assumed to be a monochromatic 
plane-wave classical field of either linear 
or circular polarization which is 
switched on and off adiabatically 
at $t \to -\infty$ and $t \to +\infty$, respectively. 
Its four-potential is given by 
\begin{eqnarray} 
A^{\nu}(x)=a_0 \left( e_1^{\nu} \cos(\varphi) + e_2^{\nu} \sin(\varphi) \right).  
\label{em-1} 
\end{eqnarray} 
In (\ref{em-1}) $a_0$ is the amplitude, $e_j^{\nu}$ 
($\nu=0,1,2,3$, $j=1,2$) with $e_1^{\nu} e_2^{\nu} =0$ 
are the polarization four-vectors 
(in the case of linear polarization $e_2^{\nu} =0$) and 
$\varphi = k_0^{\mu} x_{\mu}$ is the phase, where  
$k_0^{\mu} = (\omega_0/c; {\bf k}_0)$ 
($e_j^{\nu} k_{0,\mu} =0$) is the four-momentum 
of the laser photons having a frequency $\omega_0$ and 
a three-momentum ${\bf k}_0$ and $x_{\mu}=(ct; {\bf x})$ 
are the space-time coordinates. 
	
The transition $S$-matrix element, 
which describes the electron-atom collision 
in the laser field, can be written as 
\begin{eqnarray}
S_{fi}= - \frac{i}{c^2} \int d^4x \int d^4y \,%
J^e_{\mu}(x) \, D^{\mu \nu}(x-y) J^A_{\nu}(y). 
\label{e1}
\end{eqnarray} 
Here, $J^e_{\mu}(x)$ and $J^A_{\nu}(y)$ ($\mu, \nu=0,1,2,3$) 
are the electromagnetic transition 4-currents generated by 
the projectile-electron at a space-time point $x$ and
by the target-atom at a space-time point $y$, respectively,
and $D^{\mu \nu}(x-y)$ is the propagator 
for the electromagnetic field transmitting 
the projectile-target interaction.        
Our evaluation of the transition amplitude Eq.(\ref{e1}) 
is based on the following assumptions. 

First, the interaction between the projectile-electron 
and the target is assumed to proceed via 
a single photon exchange.  
Note that such a first order approximation 
in the electron-target interaction 
is certainly a very good one provided 
the incident (and scattered) electron 
has high (relativistic) energy and the nuclear charge 
of the target $Z_a$ is relatively low ($Z_a \ll c$).  
The high energy of the projectile-electron also 
enables one to neglect the contribution to 
the transition amplitude from the exchange diagram. 

Second, the transition 4-current $J^e_{\mu}(x)$ 
generated by the projectile-electron is calculated  
using the Gordon-Volkov states \cite{Gordon} 
which represent an exact solution for the motion 
of a free electron in the presence of an electromagnetic field 
generated by the laser. 

Third, the transition $4$-current of the target 
is evaluated with free initial and final target states.  
Such an approximation is well justified 
for bound-bound target transitions if the bound states 
are not resonantly coupled by the laser field and the latter   
is much weaker that the intra-atomic field in these states. 
Moreover, such an approximation 
can be also used to estimate the {\it total} cross 
section for the direct transitions 
to the target continuum \cite{f-phys}.  

Starting with Eq. (\ref{e1}) and using 
the above three assumptions  
one can show that the amplitude for the field-assisted 
electron-target collision, in which 
the projectile-electron scatters from the state 
$\Psi_i$ into the state $\Psi_f$ 
(see Eq.(\ref{electron-1}))
and the target undergoes a transition 
between its internal states $\psi_i$ and $\psi_f$,   
is given by    
\begin{eqnarray} 
S_{fi} = && i \frac{8 \pi^2 }{V_e} %
\lim_{\eta \to +0} \sum_{n=-\infty}^{\infty} %
\frac{ \overline{u}_f \, \Gamma_{\mu}^{(n)} \, u_i \, F_{fi}^{\mu} } %
{Q_n^2 - \frac{ \omega_{fi}^2 }{c^2} + i \eta}  %
\nonumber \\ 
&& \times \delta\left( E_i - E_f  - \omega_{fi} + n \omega_0 + 
\lambda \left( \frac{F_0}{\omega_0} \right)^2 \right),   
\label{e2}
\end{eqnarray} 
where 
\begin{eqnarray}
F_{fi}^{\mu} &=& \langle \varphi_f \mid %
Z_a - \sum_{j=1}^{N_a} e^{i {\bf Q}_n %
\cdot {\bf r}_j} \gamma^0_{(j)} \gamma^{\mu}_{(j)} \mid \varphi_i \rangle 
\label{form-factor}
\end{eqnarray} 
is the four-dimensional form-factor of the target.  
The sums over $j$ and $n$ in (\ref{e2}) indicate 
the summations over the target electrons 
(for an ion $N_a \neq Z_a$)  
and the number of the laser photons exchanged 
between the projectile-target subsystem and the field. 
Further, $\gamma^{\mu}_{(j)}$ are 
the $\gamma$-matrices for the $j$-th target electron, 
$\omega_{fi} = \epsilon_f - \epsilon_i $ 
is the change in the internal target energy, 
$V_e$ is the normalization volume for the projectile-electron 
and the delta-function expresses the energy balance in the collision. 
The momentum transfers to the atom, ${\bf Q}_n$, 
are given by  
\begin{eqnarray}
{\bf Q}_n = {\bf P}_i - {\bf P}_f + n {\bf k}_0 - 
\lambda \, \frac{F_0^2}{\omega_0^3} \, {\bf k}_0, 
\label{e3}
\end{eqnarray} 
where $F_0$ is the electric field amplitude 
of the laser field and 
\begin{eqnarray}
\lambda &=& \lambda_0 \omega_0  %
\left( \frac{1}{ k_0^{\nu} p_{i,\nu} }  
- \frac{1}{ k_0^{\nu} p_{f,\nu} } \right)   
\label{e4-new}
\end{eqnarray} 
with $\lambda_0 = 1/2$ and $\lambda_0 = 1/4$ 
for circular and linear polarization, respectively. 

The explicit form of the expression for the quantities   
$\Gamma_{\mu}^{(n)}$ depends on the field polarization and 
is rather cumbersome. Therefore, it 
will be given here only for 
a circularly polarized field. 
Choosing the polarization vectors 
as $e_1^{\nu}=(0;{\bf e}_1)$ and 
$e_2^{\nu}=(0;{\bf e}_2)$ one obtains 
\begin{eqnarray}
\Gamma^{(n), \mu} &=& e^{i n \phi}  
\left( %
\Omega_0^{\mu} J_n(x) +  \Omega_{+}^{\mu} J_{n-1}(x) e^{-i \phi}  
\right. 
\nonumber \\ 
&& \left. + \Omega_{-}^{\mu} J_{n+1}(x) e^{i \phi} \right),  
\label{e4}
\end{eqnarray} 
where $J_n$ are the Bessel functions,  
\begin{eqnarray} 
x &=& \frac{F_0}{\omega_0} \sqrt{    
\left( \mbox{\boldmath$\kappa$} \cdot {\bf e}_1 \right)^2 + 
\left( \mbox{\boldmath$\kappa$} \cdot {\bf e}_2 \right)^2 }  
\nonumber \\ 
\mbox{\boldmath$\kappa$}&=& 
\frac{{\bf P}_i }{ k_0^{\nu} p_{i,\nu} } - 
\frac{{\bf P}_f }{ k_0^{\nu} p_{f,\nu} }  
\nonumber \\ 
\phi &=& \arctan \left( 
\frac{ \mbox{\boldmath$\kappa$} \cdot {\bf e}_2}   
{ \mbox{\boldmath$\kappa$} \cdot {\bf e}_1} \right)   
\label{e5}
\end{eqnarray} 
and 
\begin{eqnarray} 
\Omega_0^{\mu} &=& \gamma^{\mu} + 
k_0^{\mu} \frac{F_0^2 }{2 \omega_0^2 c^2} \, \, %
\frac{ \not{k}_0 }%
{ k_0^{\nu} p_{i,\nu} \, \, k_0^{\alpha} p_{f,\alpha} } 
\nonumber \\  
\Omega_{+}^{\mu} &=& a_0 \gamma^{\mu} %
\frac{ \not{e}_{+} \not{k}_0 }{2c} %
\left( \frac{1}{ k_0^{\nu} p_{i,\nu}}  - 
\frac{1}{k_0^{\nu} p_{f,\nu} }\right) 
\nonumber \\ 
&& - \frac{1}{c k_0^{\nu} p_{f,\nu}} %
\left(\not{e}_{+} k_0^{\mu} 
- \not{k}_0 {e}_{+}^{\mu}\right) 
\nonumber \\  
\Omega_{-}^{\mu} &=& a_0 \gamma^{\mu} %
\frac{ \not{e}_{-} \not{k}_0 }{2c} %
\left( \frac{1}{ k_0^{\nu} p_{i,\nu}}  - 
\frac{1}{k_0^{\nu} p_{f,\nu} }\right) 
\nonumber \\ 
&& - \frac{1}{c k_0^{\nu} p_{f,\nu}} %
\left(\not{e}_{-} k_0^{\mu} 
- \not{k}_0 {e}_{-}^{\mu}\right) 
\label{e8} 
\end{eqnarray} 
with $e_{\pm}^{\mu} = 0.5 \left( e_{1}^{\mu} \pm i e_{2}^{\mu} \right) $ 
and the Feynman's slash notation $\not b$ used to denote 
the four-scalar product $\gamma_{\nu} b^{\nu}$.  

Using Eq.(\ref{e2}) the cross section 
for the target transition 
$\psi_i \to \psi_f$, differential 
in the final momentum 
of the projectile-electron,  
is obtained to be 
\begin{eqnarray}
\frac{d^3 \sigma_{fi}}{d {\bf P}_f^3} &=& \frac{4}{v_i} %
\lim_{\eta \to +0}  \sum_{n=-\infty}^{\infty} 
\frac{ \left| %
\overline{u}_f \, \Gamma_{\mu}^{(n)} \, u_i \, F_{fi}^{\mu} %
\right|^2 }%
{ \left( Q_n^2 - \frac{ \omega_{fi}^2 }{c^2} \right)^2 + \eta^2 }    %
\nonumber \\ 
&& \times \delta\left( E_i - E_f  - \omega_{fi} + n \omega_0 - 
\lambda \left( \frac{F_0}{\omega_0} \right)^2 \right), %
\label{e9}
\end{eqnarray} 
where $v_i$ is the initial velocity of the incident electron. 
If the initial state of the projectile-electron 
is unpolarized and its final polarization 
is not detected the above cross section 
should be averaged/summed over the electron spin.  

Note that the cross section (\ref{e9}) is valid both 
for the elastic and inelastic (for the target) collisions. 
In the latter case the interaction between 
the projectile-electron 
and the target nucleus, whose contribution 
to the amplitude (\ref{e2}) is proportional 
to $Z_a$, does not influence the collision process.  

In the limit $F_0 \to 0$ one has 
$\Gamma^{(n),\mu} \to \delta_{n0} \gamma^{\mu}$  
and formula (\ref{e9}) reduces to the well known expression 
for the (first order) cross section describing 
the field-free collision between a relativistic electron 
and an atom. Besides, one can also prove that 
if the incident electron is not relativistic, 
the laser field is not too strong and 
the initial and final states of the target 
can be well approximated by non-relativistic 
(Schroedinger) wave functions,    
the cross section (\ref{e9}) goes over into 
the expression well known from 
the non-relativistic theory.  

The relativistic effects 
in the cross section (\ref{e9})  
can be characterized as: 
(i) those depending on 
the initial kinetic energy of the incident 
electron, (ii) those related to the parameters 
of the laser field and (iii) 
those connected with relativistic effects 
in the motion of the target electrons 
in their initial and final states. 

Concerning (i) and (ii) note that 
noticeable deviations from the non-relativistic 
description of the collision process 
will obviously be present 
if the initial (at $t \to - \infty$) velocity 
of the incident electrons is comparable to $c$. 
Moreover, even if this velocity 
is much less than $c$, the laser field can, 
provided $F_0/\omega_0 \stackrel{>}{\sim} c$, 
accelerate the incident electrons  
to relativistic energies making it necessary 
to describe the collision process relativistically. 

\begin{figure}[t]  
\vspace{-0.25cm}
\begin{center}
\includegraphics[width=0.33\textwidth]{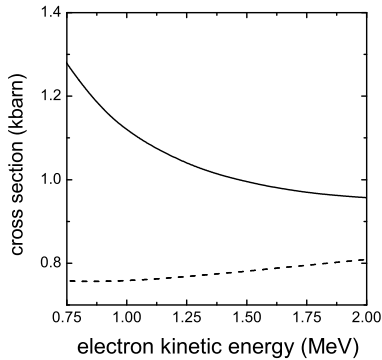}
\end{center}
\vspace{-0.75cm} 
\caption{ \footnotesize{ 
Ne$^{9+}$(1s) $\to $ Ne$^{9+}$(n=2) 
transitions caused by electron impact. 
(a) Dash curve: field-free case. 
Solid curve: collisions assisted by a circularly 
polarized laser field with $F_0=2$ a.u. 
and $\omega_0 = 4.3 \times 10^{-3}$ a.u.  
for the head-on beam collision geometry  
(see the text). } } 
\label{figure1}
\end{figure}

Let us now consider two simple examples 
of the application of the theory developed above 
in which the presence of the laser field influences 
target transitions via qualitatively different 
reaction pathways.  

1. In figure \ref{figure1} we show results for excitation 
of Ne$^{9+}(1s)$ ions into bound states with 
the principal quantum number $n=2$ caused 
by collisions with electrons.  
The excitation cross section are shown 
for the field-free and field-assisted collisions. 
In the latter case the laser field is assumed 
to be circularly polarized and propagate 
in the direction opposite to the initial 
momentum of the incident electrons 
(the head-on beam collision geometry).  
The strength and frequency of the field are 
$F_0=2$ a.u. (the corresponding field intensity 
is $\simeq 1.4 \times 10^{17}$ W/cm$^2$) 
and $\omega_0 =4.3 \times 10^{-3}$ a.u. ($0.117$ eV).  

This laser field is much 
weaker then the typical fields in the 
initial and final states of the Ne$^{9+}$ ion. 
Therefore, the laser field does not  
influence the transitions directly. 
However, since $F_0/\omega_0 > c$, 
the laser field can very substantially 
affect the motion of the incident electrons. For instance, 
the change in the kinetic energy of these electrons 
due to their interaction with the field can 
be comparable to their initial kinetic energy which 
substantially modify the dynamics 
of the electron-ion collision.  
 
As one can expect in relativistic collisions 
the magnetic component of the laser field and its momentum 
start to have an important impact 
on the target transitions. 
In particular, in the example 
under consideration 
the influence of the momentum 
of the photons exchanged between the field 
and the colliding electron-target system 
on the collision dynamics becomes comparable to that 
of the energy of these photons.   

2. The relativistic treatment describes 
one more interesting physical effect 
which is absent in the non-relativistic consideration. 
In the relativistic description the denominator in the 
transition amplitude (\ref{e2}) contains, 
compared to its non-relativistic counterpart, 
extra terms. In addition to the terms related 
to the momenta of the laser photons  
(and to the fact that in the relativistic case 
the ponderomotive energies in the initial and final states of 
the incident electrons do not cancel each other), 
in the denominator there appears 
the so called retardation term, 
$\omega_{fi}^2/c^2$, which is 
the square of the 'time' component 
of the four-momentum 
transfer ($\omega_{fi}/c, {\bf Q}_n$) 
between the incident electron and the target.  
The presence of the retardation term may 
lead to a singularity \cite{f1}.  

The denominator in Eq. (\ref{e2}) becomes singular 
when the interaction between the incident 
electron and the target is transmitted 
by on-mass-shell photon that dramatically  
increases the effective range  
of the electron-target interaction.    
The appearance of the singularity is directly related 
to the fact that a relativistic electron moving 
in a laser field can under certain conditions 
convert laser photon(s) with a frequency $\omega_0$ 
into a photon whose frequency $\omega$ is resonant 
to a certain target transition. The physical mechanism 
for this conversion is the Compton effect.       

\begin{figure}[t]  
\vspace{-0.25cm}
\begin{center}
\includegraphics[width=0.33\textwidth]{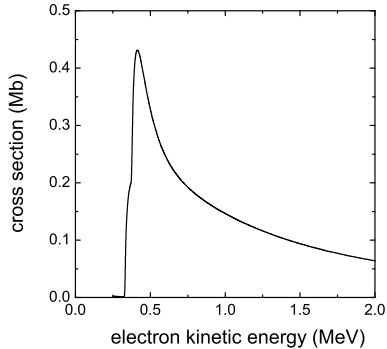}
\end{center}
\vspace{-0.75cm}
\caption{ \footnotesize{ H(1s) $\to$ H(n=2) 
transitions caused by the interaction 
with the counter propagating beams of 
relativistic electrons and laser radiation.  
The laser field is linearly polarized, 
$F_0=0.5$ a.u. and $\omega_0= 4.3 \times 10^{-2}$ a.u.. 
See the text for more explanations. } }  
\label{figure2}
\end{figure} 
In such a case the electron-target interaction 
can be thought of as occurring via two steps: 
(i) the emission of a resonant photon by 
the incident electron and ii) the absorption of this 
photon by the target. 
In general the cross section (\ref{e9}) 
does not enable one to describe 
the two-step process. The standard way 
of handling such a situation is    
to remove the singularity from 
the cross section by an appropriate 
regularization procedure  
and to evaluate the number 
of physical events corresponding 
to the two-step process by 
computing, for a given geometry of 
the electron and laser beams, 
the momentum distribution 
of the resonant photons considered 
as a function of space-time (and 
the properties of the target medium)  
and then to use this distribution 
to estimate the probabilities 
for the target transitions caused by 
absorption of these photons.  
 
However, for especially simple 
collision geometries, one can still 
try to get a rough estimate of the number 
of total physical events 
based on the cross section (\ref{e9}). 
This can be done by taking into account 
the target density and size effects \cite{we-res} 
that enable one to regularize (\ref{e9}). 
The resulting effective cross section 
is related to the number 
of physical events like the normal cross section.  

In figure \ref{figure2} we show 
estimates for the cross section 
for the H(1s) $\to$ H(n=2) 
transitions caused by the interaction 
with colliding beams of 
relativistic electrons and laser radiation. 
The laser field is linearly polarized,  
$F_0=0.5$ a.u. and $\omega_0= 4.3 \times 10^{-2}$ a.u..  
We assume that the target gas with 
a rather low density  
($ < 10^{12}$ cm$^{-3}$) is contained 
in a volume whose boundaries are 
coaxial cylinders with the radii of $0.1$ 
and $0.5$ mm, respectively, 
and the length of $2$ mm.    
Counter propagating 
beams of electrons and laser field 
with equal diameters of $0.1$ mm penetrate 
this space along its symmetry axis. 

Since these beams do not penetrate 
the volume occupied by the target gas, 
none of them taken separately 
would induce transitions in the target.  
However, when both beams 
are simultaneously present,  
the Compton scattering and the  
corresponding two-step process 
come into play leading 
to the target excitation. 
Note also that in this example,     
since $F_0/\omega_0 \ll c$, 
the laser field does not  
change substantially the motion of 
the incident electrons.  

The cross section in figure 
\ref{figure2} can be compared 
with the cross section for the H(1s) $\to$ H(n=2) 
excitation by relativistic electrons in case when they 
penetrate the target medium (the laser is absent) 
which varies approximately between  
$0.15$ and $0.2$ Mb for the impact energy 
$0.25$ - $2$ MeV.  
 
In conclusion, briefly, we have considered 
inelastic collisions 
of relativistic electrons with 
atomic targets assisted by a laser field. 
The parameters of the field were assumed 
to be such that the field does not have a 
noticeable direct impact on the 
target transitions and can influence them 
only via its interaction with the incident electrons.   

We have shown that the presence 
of such a field can substantially 
change inelastic target cross sections 
in two qualitatively different situations.  
First, laser fields with the strength 
and frequency satisfying the condition 
$F_0/\omega_0 \stackrel{>}{\sim} c$ 
influence target transitions by inducing 
strong modifications in the motion 
of the relativistic electrons. 
Second, relatively weak laser fields,  
$F_0/\omega_0 \sim  1$,  may 
impact target transitions via the Compton 
effect when relativistic electrons transform 
laser radiation into photons whose frequencies 
are resonant to target transitions.


\begin{thebibliography}{99}
\bibitem{f-phys} 
The transfer of a bound electron 
into the target continuum, 
caused by the impact of a fast particle,   
effectively occurs during 
the very short transition time 
$\tau \sim 1/\omega_{fi}$, where 
$\omega_{fi}$ is the transition frequency. 
The modifications of the target continuum 
by a low-frequency laser field 
are accumulated on the much longer 
time scale $T \sim 1/\omega_0 \gg \tau$.
Therefore, the laser field with 
$F_0 \ll F_a $, where $F_a$ 
is the typical atomic field 
in the initial state, 
is not really able  
to affect the transition 
on the time scale $\sim \tau$. 
As a result, the field 
cannot noticeably change   
the total amount of 
target electrons transferred 
to the continuum by the projectile-electron  
impact (although can strongly modify, 
compared to the field-free case,  
the spectra of the emitted electrons 
by interacting with 
them on the much longer time scale).
\bibitem{f1} Note that in the non-relativistic description 
the 'time' component of the four-momentum transfer 
does not appear and the singularity in the photon  
propagator can arise only if ${\bf Q}_n=0$. 
However, since at ${\bf Q}_n=0$  
the inelastic target form-factors  
are equal to zero, target transitions 
do not occur.
\end{thebibliography}
\end{document}